# Negative reflection and negative refraction in biaxial van der Waals materials

*Tan Zhang,[1,†] Chunqi Zheng,[1,2,†] Zhi Ning Chen,[1] and Cheng-Wei Qiu[1,*]*

[1]Department of Electrical and Computer Engineering, National University of Singapore, Singapore 117583, Singapore

[2]NUS Graduate School, National University of Singapore, Singapore 119077, Singapore

**ABSTRACT**

Negative reflection and negative refraction are exotic phenomena that can be achieved by platforms such as double-negative metamaterial, hyperbolic metamaterial, and phase-discontinuity metasurface. Recently, natural biaxial van der Waals (vdW) materials, which support extremely anisotropic, low-loss, and highly confined polaritons from infrared to visible regime, are emerging as promising candidates for planar reflective and refractive optics. Here, we introduce three degrees of freedom, namely interface, crystal direction, and electric tunability to manipulate the reflection and refraction of the polaritons. With broken in-plane symmetry contributed by the interface and crystal direction, distinguished reflection and refraction such as negative and backward reflection, positive and negative refraction could exist simultaneously and exhibit high tunability. The numerical simulations show good consistency with the theoretical analysis. Our findings provide a robust recipe for the realization of negative reflection and refraction in biaxial vdW materials, paving the way for the polaritonics and interface nano-optics.

**INTRODUCTION**

Since Victor Veselago[1] first proposed the double-negative (permittivity $\varepsilon < 0$, permeability $\mu < 0$) medium in 1968, considerable milestone research[2–5] has been conducted to provide consolidated theoretical support for the exotic phenomenon—negative refraction. Classic artificial structures such as split-ring resonators were used to construct the metamaterials operating in microwave regime[6–8]. Later, hyperbolic metamaterials[9] realized by multilayered[10] and fishnet-like[11] structures were utilized to achieve negative refraction in terahertz and optical frequencies with a broadened operation band. Moreover, metasurface[12,13] was proposed to realize abnormal reflection and refraction by introducing phase discontinuities along the interface, which greatly advances the development of flat optics[14,15]. Recently, natural van der Waals (vdW) materials[16–19] have attracted continuous attention and emerged as a powerful platform for the manipulation of electromagnetic waves because they can support low-loss, extreme anisotropic, and highly confined polaritons (hybrid excitation of matter and photons[20]). Planar reflection, refraction, and focusing were demonstrated by using composite systems consisting of graphene[21,22], hBN[23], and phase-changing materials[24,25]. Furthermore, biaxial vdW materials, such as alpha-molybdenum trioxide (α-MoO$_3$)[26–29], exhibit in-plane hyperbolicity within the Reststrahlen bands, acting as an additional degree of freedom that guarantees a simplified platform for planar reflection and refraction of polaritons.

In this work, by engineering the interface and rotating the crystal direction of the transmitted α-MoO$_3$, both negative reflection and negative refraction can be achieved. Additionally, the reflection and refraction can be tuned from positive to negative (Figure 1b) and the critical conditions for this transition are governed by mathematical equations deriving from fundamental geometric relations. Furthermore, with broken in-plane symmetry, the refraction of polaritons is not "binary" but can be positive and negative simultaneously, see Figure 1c. Finally, we demonstrate electrically tunable focusing by introducing naturally anisotropic material black phosphorous (BP). Our findings realize the high tunability of polaritons and serve as comprehensive guidance for the realization of in-plane negative reflection and negative refraction in biaxial vdW materials, advancing the development of interface nano-optics.

The major principle used here is the well-known phase-matching conditions[30], which states that the tangential component of the wavevectors along the interface should be continuous ($\mathbf{k}_\mathbf{i}^\parallel = \mathbf{k}_\mathbf{r}^\parallel = \mathbf{k}_\mathbf{t}^\parallel$, Figure 1a). Then, we could sketch the incident, reflected, and refracted wavevectors and determine the reflection and refraction of the polaritons accordingly, see the left panel of Figure 1 (find details in Section S1). Note that the reflection and refraction beam are mainly reflected/represented by the Poynting vector **S**. Therefore, negative reflection (refraction) is realized when the reflected Poynting vector $\mathbf{S_r}$ (transmitted $\mathbf{S_t}$) and the incident Poynting vector $\mathbf{S_i}$ are on the same side of the norm of the interface. Besides, for convenience, we define the incident polaritons with positive amplitudes of Poynting vectors along [100] and [001] directions ($S_{100}$, $S_{001}$ > 0) as upper-polaritons (UPs), and the ones with positive amplitudes along [100] and negative amplitudes along [001] ($S_{100}$ > 0, $S_{001}$ < 0) as lower-polaritons (LPs). In this circumstance, all the analysis and artistic illustrations in Figure 1 are concerned with the UPs.

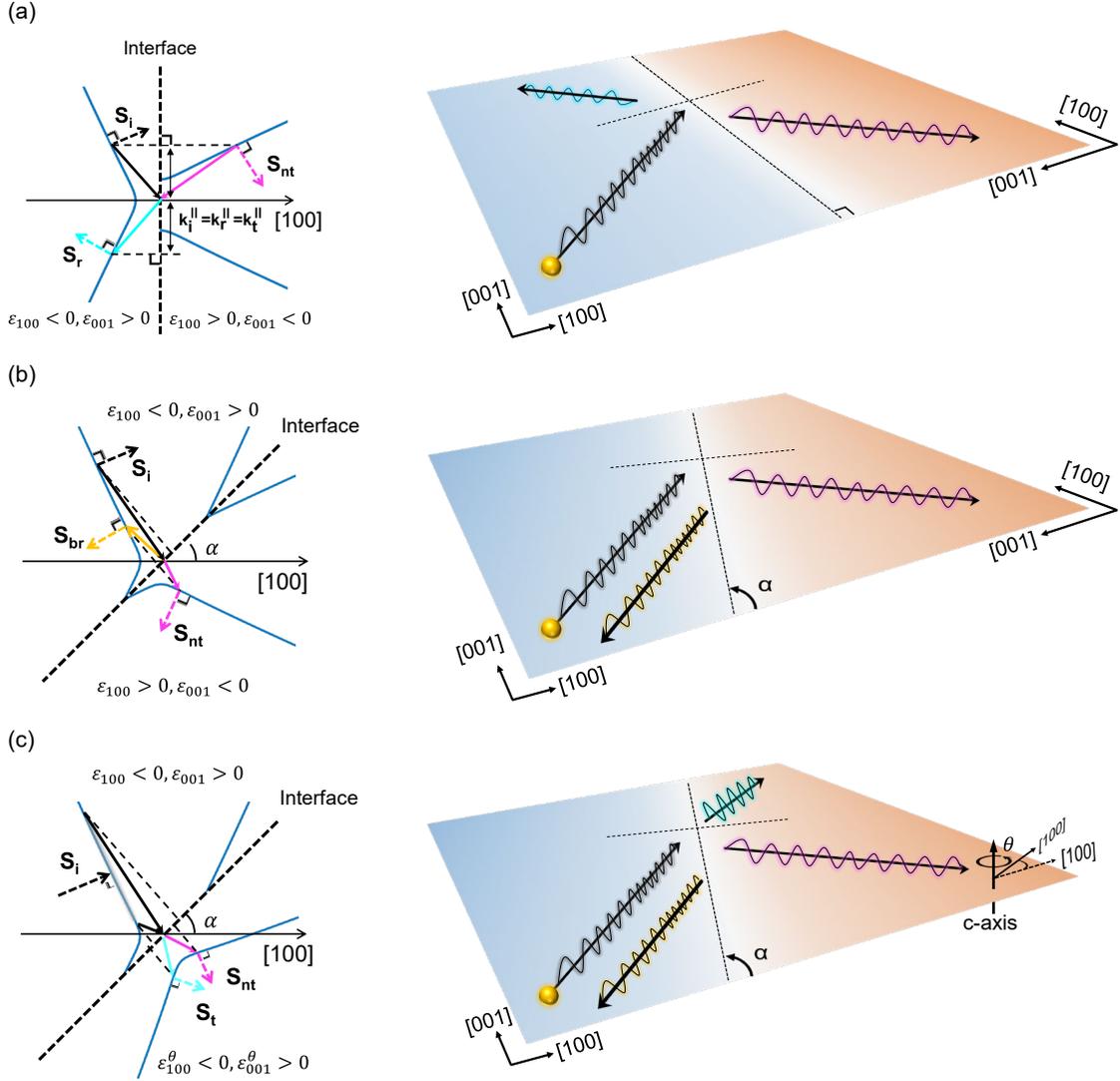

**Figure 1.** Schematics of negative reflection and negative refraction. (a-c) The reflection and refraction of polaritons with varying interface angle $\alpha$ and rotation angle $\theta$ in k-space (left), and their corresponding artistic illustrations in real space (right). The square-dotted line is the interface, and the dashed line is the norm of the interface. The blue curves on the right- (left-) side of the interface stand for the dispersion of the incident (transmitted) medium. The second convention is used for demonstrating reflection and refraction (see Section S1). (a) Negative refraction $S_{nt}$ and positive reflection $S_r$ with a perpendicular interface ($\alpha = 90°$) and a transmitted medium with 90°-rotated dispersion ($\theta = 90°$). $\varepsilon_{100}$ ($\varepsilon_{001}$) is the permittivity along the rotated [100] ([001]) crystal direction. (b) Negative refraction $S_{nt}$ and negative reflection (specifically backward reflection $S_{br}$) with a titled interface ($|\alpha| < 90°$) and a 90°-rotated dispersion. (c) Negative $S_{nt}$ and positive $S_t$ refraction coexist with a titled interface $|\alpha| < 90°$ and a rotated dispersion ($|\theta| < 90°$). $\varepsilon^{\theta}_{100}$ ($\varepsilon^{\theta}_{001}$) is the permittivity along the rotated [100] ([001]) crystal direction.

## RESULTS

**Negative reflection**

The natural low-loss biaxial vdW material α-MoO$_3$ with a tilted interface between a reflector (e.g., air) is used to demonstrate the negative reflection. $\alpha$ ($\in$ [-90°, 90°], positive for the anticlockwise direction) is the angle between the tilted interface and the [100] crystal direction (Figure 1b). It is found that the transitions between positive and negative reflection are governed by certain critical angles, and the range where negative reflection happens is given as follows:

$$|\alpha| \in [90° - \beta, \beta] \qquad (1)$$

where $\beta = \mathrm{atan}\sqrt{-\varepsilon_{100}/\varepsilon_{001}}$ is the open-angle of the dispersion curve (assuming $\beta > 45°$), $\varepsilon_{100}$ and $\varepsilon_{001}$ are the permittivity along the [100] and [001] crystal direction respectively. A 200 nm α-MoO$_3$ flake illuminated by light at wavenumber $\omega$ = 900.3 cm$^{-1}$ (open-angle $\beta$ = 66°) is used for demonstration. The reflection can be determined by choosing three representative incident wavevectors **k** and their corresponding Poynting vectors **S** (see Figure 2b): **k**$_{i,\,\mathrm{I}}$ and **S**$_{i,\,\mathrm{I}}$; **k**$_{i,\,\mathrm{II}}$ and **S**$_{i,\,\mathrm{II}}$; **k**$_{i,\,\mathrm{III}}$ and **S**$_{i,\,\mathrm{III}}$. (see details in Section S2)

Figure 2b shows the complete analysis of the reflection at the interface angle $\alpha$ = 45°. The simplified schematics of the negative reflections at $\alpha$ = 45° are summarized in the middle panel of Figure 2d. The broken in-plane symmetry introduced by the tilted interface leads to distinguished refraction of the polaritons: backward reflection (II) for the UPs and negative reflections (III) for the LPs. The simulated electric field distribution of $|E_z|$ visualizes the backward reflection **S**$_{br}$ and the negative reflection **S**$_{nr}$ (Figure 2f). Note that, the negative reflection (I) of the UPs contributed by the first representative wavevector **k**$_{i,\,\mathrm{I}}$ is hardly seen in the simulated results, which could be explained by two reasons: one is that for the UPs, the coverage of the backward reflection region (yellow) is larger than that of the negative reflection (pink), see Figure 2b; the other is that the negative reflection (I) of the UPs is contributed by polaritons with larger wavenumbers ($|\mathbf{k}_{i,\,\mathrm{I}}| > |\mathbf{k}_{i,\,\mathrm{II}}|$), which are hardly excited and exhibit large loss. Therefore, the negative reflection **S**$_{nr,\,\mathrm{I}}$ becomes trivial while the backward reflection **S**$_{br,\,\mathrm{II}}$ prevails.

The complete analysis of the case where $\alpha$ = 45° shows that the reflection of UPs or LPs can be determined by a single wavevector without the loss of generality. Figure 2a and 2c exhibit the theoretical reflections of UPs at four critical interface angles. At $\alpha$ = ±24° (90° - $\beta$), both the norms (black dashed line) of the interfaces are parallel to the asymptote of the dispersion, therefore there's no solution for the reflected wavevectors, and the reflection is prohibited at these interface angles. While for $\alpha$ = ±66° ($\beta$), it is the interface (red dashed line) that is parallel to the dispersion curve, therefore either the reflected or incident Poynting vectors are normal to the interface. In this circumstance, we defined it as critical reflection considering neither positive nor negative reflection is appropriate. By further increasing $|\alpha|$, the reflected and incident Poynting vectors will be tuned to the opposite side of the norm, bringing about positive reflection. Hence, when $|\alpha| \in [90° - \beta, \beta]$ (e.g., 45°), the polaritons would be negatively reflected and the simulated field distributions of UPs presented in Figure 2e-g keep good consistence with the analysis above.

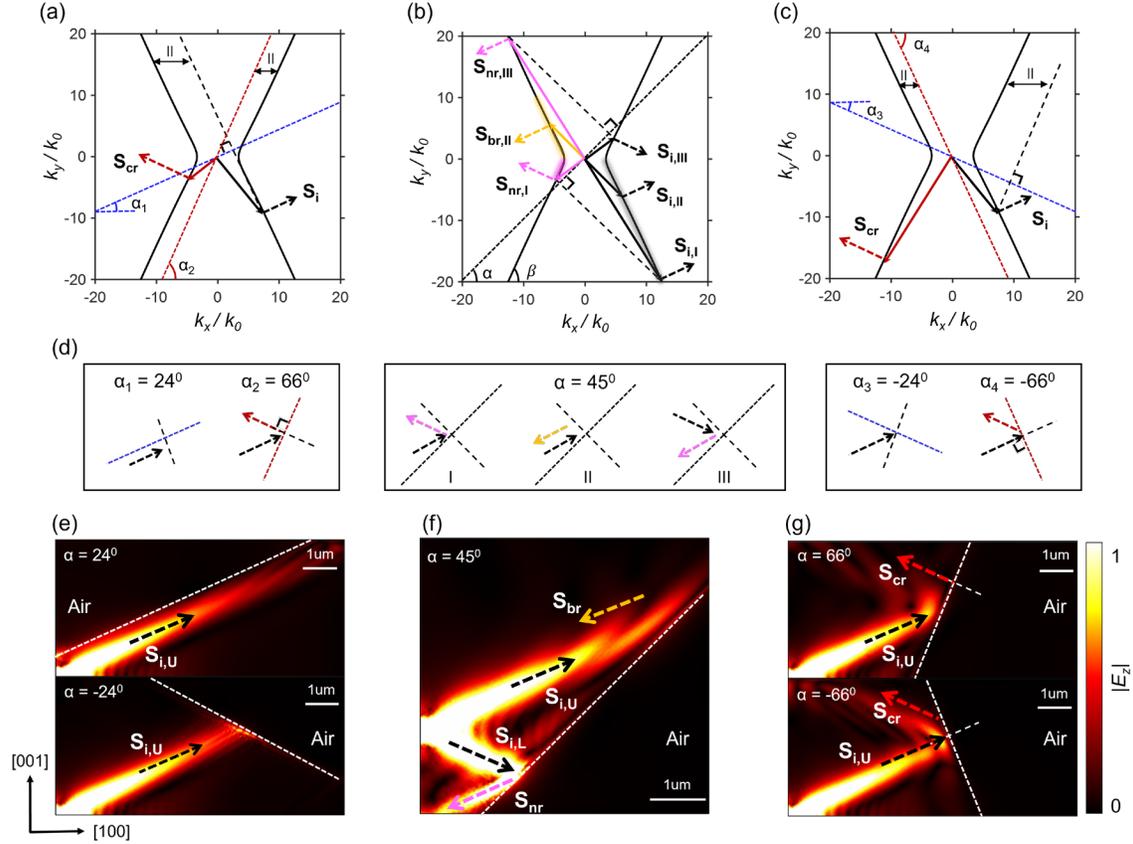

**Figure 2.** Interface angle $\alpha$ dependence of negative reflection. (a-c) The theoretical analysis of the reflection of polaritons at five different interface angles: (a) $\alpha = 24°, 66°$; (b) $\alpha = 45°$; (c) $\alpha = -24°, -66°$. The solid arrows represent the wavevectors **k** and the dashed arrows indicate their corresponding Poynting vectors **S**. The abbreviations in the subscripts of **S** such "i", "r", "nr", and "cr" stand for incidence, backward reflection, negative reflection, and critical reflection, respectively. (d) The simplified schematics of the incident and refracted **S**. (e-g) The numerically simulated field distribution of the norm of the z-component of electric field ($|E_z|$) at different interfaces angles: $\alpha = \pm 24°, 45°$, and $\pm 66°$. $S_{i, U}$ ($S_{i, L}$) represents the major energy flow of the upper (lower) polaritons. The Poynting vectors in (e-g) are strictly parallel to that in (a-d).

## Negative refraction

The refraction in biaxial vdW material manifests a much more exotic behavior, where positive and negative refraction can happen simultaneously. In addition, not only the interface but also the transmitted medium would impose significant influences on the refraction of the polaritons. Here, we replaced the perfect reflector by α-MoO$_3$ with a rotating crystal direction $\theta$ ($\in$ [-90°, 90°], positive for anticlockwise direction) to study the in-plane refraction, where $\theta$ is defined as the angle between the [100] crystal directions of the transmitted and incident α-MoO$_3$ (see Figure 1c). It is found that there also exist critical conditions that determine the transitions between positive and negative refraction, where $\alpha$ and $\theta$ satisfy the relations given as:

$$|\theta - \alpha| = \beta$$
$$|\theta - \alpha - 90°| = \beta \quad (2)$$

The parameters of α-MoO$_3$ used here are the same as that in reflection so that the open-angle remains unchanged ($\beta$ = 66°). For the solutions of Eq. 2 exceeding [-90°, 90°], we could modify them by $\pm$180° to make sure they stay within this range (see Section S6). Then, to have a better understanding of Eq. 2, we first fix the interface angle $\alpha$ = 45° and study the $\theta$-dependence of refraction. Figure 3a-f provide three typical cases of refraction at the negative range of $\theta$ ([-90°, 0]), including two critical angles calculated by Eq. 2 (-21° and -69°) and one transition angle between them (e.g., -45°), see the positive range of $\theta$ in Section S3. When $\theta$ = -21°, all the UPs are positively refracted, see the dashed cyan arrows $S_{t, I}$ and $S_{t, II}$ in Figure 3a, d. As for the LPs, the refracted direction $S_{ct, III}$ (red dashed arrow) is normal to the interface and classified as critical refraction, thus serving as the boundary between positive and negative refraction. By further decreasing $\theta$ (Figure 3b, e, $\theta$ = -45°), the incident and refracted Poynting vectors of the LPs will be tuned to the same side of the norm, bringing about the negative refraction (pink arrow). Meanwhile, the UPs represented by $k_{i, I}$ remain to be positively refracted (cyan arrow). In this circumstance, the non-trivial phenomena, where both negative and positive refraction happen simultaneously, is observed. When $\theta$ decreases to the second critical angle -69°, all the polaritons will be negatively reflected, see Figure 3c, f.

Secondly, we fix the crystal direction $\theta$ = 90°, and study the $\alpha$-dependence of refraction. $\theta$ = $\pm$90° is a special case when applying Eq. 2, because the angle between the transmitted and incident Poynting vectors is always equal to or greater than 90°, which means the transmitted beam would only be negatively refracted. However, the critical conditions for $\alpha$ still work and govern the refraction of the polaritons. Figure 3g-l provide three typical cases of refraction at the positive range of $\alpha$ ([0, 90°]), including two critical angles calculated by Eq. 2 (24° and 66°) and one transition angle between them (e.g., 45°), see the summary of $\alpha$-dependence in Section S5. When $\alpha$ = 66°, the UPs are negatively refracted while the transmitted LPs propagate along the interface, because the only solution $S_{nt, II}$ allowed for LPs is parallel to the interface (Figure 3g, j). With a further decrease of $\alpha$ (Figure 3h, 45°), we notice that two refraction modes with different Poynting vectors ($S_{nt, I}$ and $S_{nt, II}$) can be generated from the UPs with the same incident angle ($S_{i, I}$ and $S_{i, II}$). This "double refraction" is also indicated in the study of $\theta$-dependence, see $S_{t, I}$ and $S_{ct, II}$ in Figure 3b, e. Note that the refraction beam $S_{nt, I}$ is very weak and almost invisible in the simulation results, due to the same reasons (coverage region and loss) discussed in *Negative Reflection*. As $\alpha$ decreases to 24°, another critical condition is reached, where the interface is parallel to the asymptote of the transmitted dispersion. In this case, the UPs do not interact with the interface and only the LPs are negatively refracted (see Figure 3i, l). The Fourier transform of the simulated real part of the $z$-component of the electric field [Re ($E_z$)] is conducted and presented as the background of Figure 3a-c, g-i (see Section S4), showing good consistence with the analysis performed by the analytical dispersion curve.

To summarize, we first investigate the $\theta$-dependence of the refraction at $\alpha = 45°$. Four values of $\theta$ are obtained by solving Eq. 2. Then, by arranging them in order of smallest to largest (e.g., $\theta_1 = -69°$, $\theta_2 = -21°$, $\theta_3 = 21°$, $\theta_4 = 69°$), we can determine the refraction of the polaritons in each specific range: [-90, $\theta_1$] and [$\theta_4$, 90], negative refraction; [$\theta_1$, $\theta_2$] and [$\theta_3$, $\theta_4$], simultaneously negative and positive refraction; [$\theta_2$, $\theta_3$], positive refraction (see Figure S4). Also, the special cases of fixed $\theta$ (90°) were studied to demonstrate the $\alpha$-dependence of the refraction. Although only negative refraction happens in this circumstance, the refraction behavior of the polaritons varies within different ranges of $\alpha$ (see Table S1).

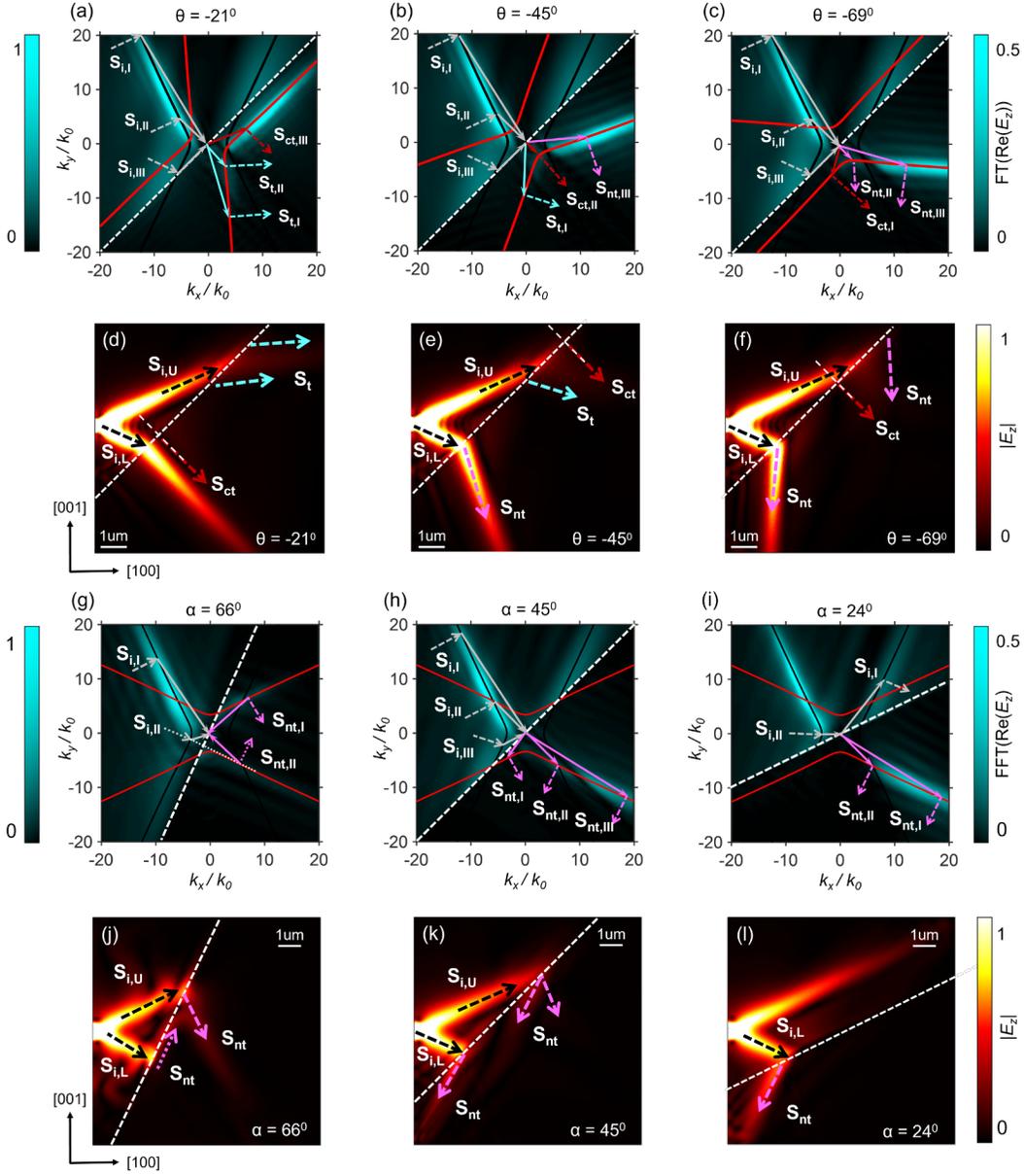

**Figure 3.** Crystal direction $\theta$ and interface angle $\alpha$ dependence of negative refraction. (a-c) The theoretical analysis and (d-f) their corresponding numerically simulated field distributions of $|E_z|$ of the refraction at three different $\theta$ (-21°, -45°, and -69°), where $\alpha$ is fixed as 45°. (g-i) The theoretical analysis and (j-l) their corresponding numerically simulated field distributions of $|E_z|$ of the refraction at three different $\alpha$ (66°, 45°, and 24°), where $\theta$ is fixed as 90°. (a-c, g-i) The black (red) curves indicate the analytical dispersion of the incident (transmitted) medium. The background graphics are the Fourier transform (FT) of the real part of the z-component of the electric field (Re ($E_z$)) in (d-f, j-i) (see details in Section S4). The left- (right-) hand sides color bars correspond to the FT in the incident (transmitted) medium. The Poynting vectors (dashed arrows) are strictly parallel to that in (d-f, j-l).

**Electrically tunability**

Focusing is an intuitive function that can be achieved through negative refraction[31,32]. Take one simple case with $\theta = 90°$, one peculiar property of focusing around $|\alpha| = 90°$ is that the refracted direction is independent of $\alpha$. Hence, by slightly rotating $\alpha$ around point o (the midpoint of the interface), the focal point can be shifted laterally ([001]) with a small moving distance in the axial ([100]) direction, see Figure 4a-c where $\theta = 90°$. The lateral shift $\Delta$ can be expressed by $\alpha$ and $\beta$ as follows

$$\Delta = -\frac{d}{2}\frac{\sin\alpha}{\cos\beta}\left(\frac{1}{\cos(\alpha+\beta)} + \frac{1}{\cos(\alpha-\beta)}\right) \tag{3}$$

where $d$ is the distance between the source and point o. In Eq. 3, $\Delta > 0$ ($< 0$) means that the focal point moves upward (downward). Field distributions $|E_z|$ with $\alpha = 75°, 90°, -75°$ (105°) are shown in Figure 4a-c, respectively, where the movement of the focal point in the lateral direction is observed. With $|\alpha| = 75°$, $\beta = 66°$ and $d = 5$ μm, the theoretical values (~1.6 μm) show good agreement with the simulated results in Figure 4a-c.

When $\alpha = 90°$, the incidence and refraction beams of UPs and LPs are symmetrical, and the refraction beams are normal to the incident ones due to the 90° rotation of the material on the transmitted side. Then, the focal length $f$ can be approximated by $d/\tan^2\beta$. Hence, it is possible to tune the focal length with the same object distance by modulating the open-angle $\beta$. Without the loss of generality, we study the thin layer of BP, a kind of vdW material with natural in-plane anisotropy arising from its peculiar orthorhombic lattice with two distinguishable axes—armchair (AC) and zigzag (ZZ) direction. Similar to the setup in α-MoO$_3$, we place the BP layer with two orthogonal directions in the incident (AC along x-direction) and the transmitted medium (AC along y-direction). Its electrical and optical properties can be modulated via thickness control[33], doping[34], and external gate bias[35]. By adjusting the gate voltage, the Hall coefficient, conductance, and carrier concentration can be modulated[36–38]. As the surface conductivity of BP relies on the concentration of electrons, we demonstrate that focus adjustment can be achieved via electric tuning. Figure 4d-f show the $|E_z|$ distribution with different electron concentrations $n$. It is shown that the focal point moves towards the interface as $n$ increases due to the increase of the open-angle (Figure 4g). The functions of open-angle $\beta$ versus electron concentration $n$ with $\omega = 1900, 2000, 2100$ cm$^{-1}$ are shown in Figure 4h. We also notice that as operation frequency $\omega$ increases, the open-angle $\beta$ will decrease (Figure 4h), leading to the increasement of focal length $f$.

Based on the regulation of the open-angle through external bias, the same focal length under different operation frequencies can be achieved by adjusting $n$. Here we choose $\beta = 72.3°$ at $1.0\times10^{14}$ cm$^{-2}$, 1900 cm$^{-1}$ ($n, \omega$) as a reference and adjust electron concentration as $1.14\times10^{14}$ cm$^{-2}$ at 2000 cm$^{-1}$, $1.30\times10^{14}$ cm$^{-2}$ at 2100 cm$^{-1}$ to keep open-angle $\beta$ unchanged. Figure 4i shows the electric field $|E_z|$ along the normal direction of the interface. It is well visible that the three transmitted beams focus on the same position in the axial direction,

showing that the fixed focal points under different frequencies can be realized through electric tuning without modifying the physical setup.

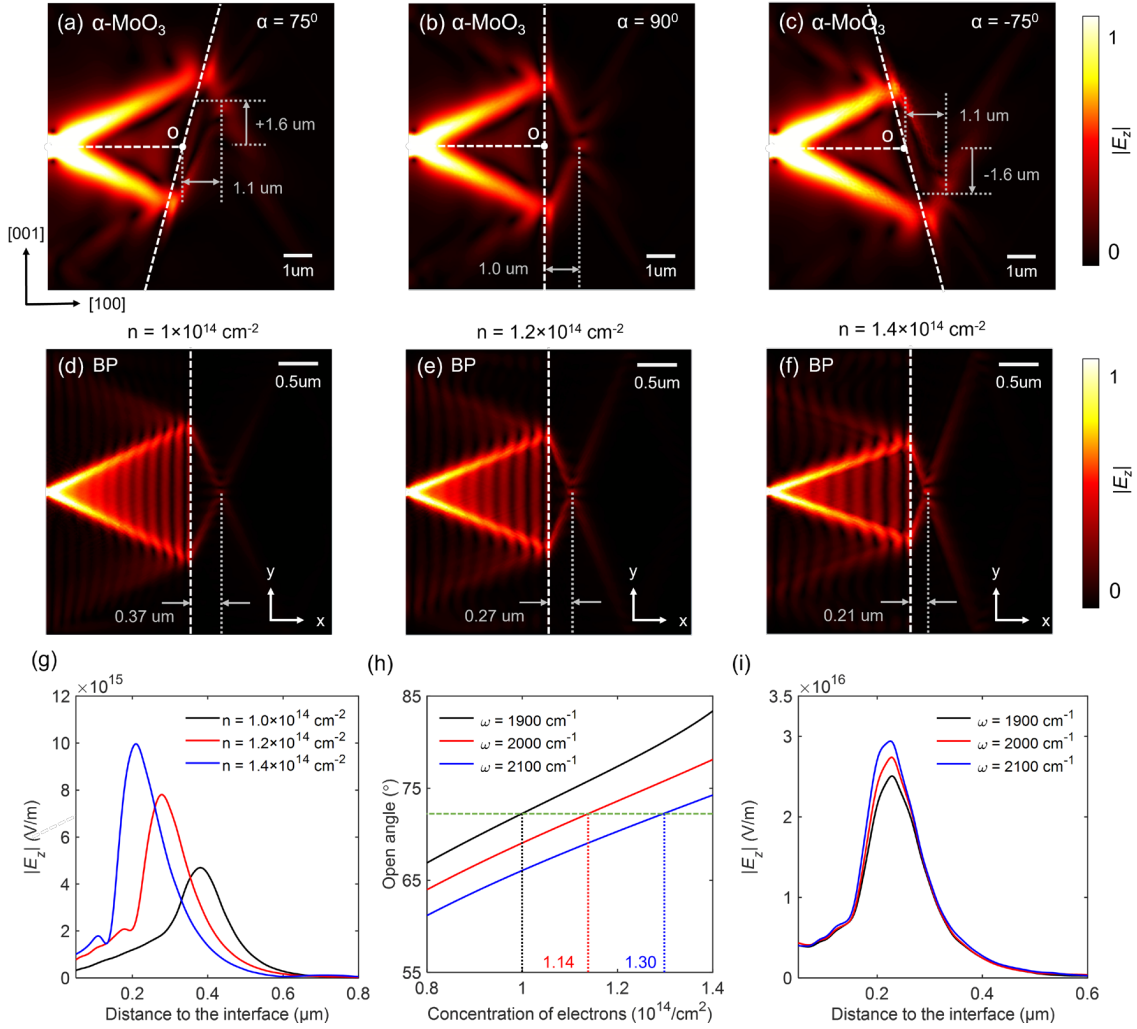

**Figure 4.** Tunable focusing. (a-c) The numerically simulated field distribution of $|E_z|$ for which $\alpha$ = 75°, 90°, and -75°, where $\theta$ is fixed as 90°. Focal point adjustment along lateral direction is shown with small moving distance along axial direction. (d-f) The numerically simulated field distribution of $|E_z|$ of BP thin layer for which $n = 1.0 \times 10^{14}$ cm$^{-2}$, $1.2 \times 10^{14}$ cm$^{-2}$ and $1.4 \times 10^{14}$ cm$^{-2}$, where $\theta = 90°$ and the operation frequency $\omega = 2000$ cm$^{-1}$. (g) $|E_z|$ changes along x-direction with y = 0. (h) The function of open angle versus concentration of electrons at three different $\omega$: 1900 cm$^{-1}$, 2000 cm$^{-1}$, and 2100 cm$^{-1}$. (i) $|E_z|$ changes along x direction with y = 0 of three cases of $n$ - $\omega$: $1.0 \times 10^{14}$ cm$^{-2}$ - 1900 cm$^{-1}$, $1.14 \times 10^{14}$ cm$^{-2}$ - 2000 cm$^{-1}$ and $1.30 \times 10^{14}$ cm$^{-2}$ - 2100 cm$^{-1}$.

## CONCLUSIONS

In this work, we first demonstrate the realization of negative reflection at a tilted interface of biaxial vdW material, where negative reflection happens when $|\alpha| \in [90° - \beta, \beta]$. Moreover, backward reflection is observed for $\alpha$ within this range. Following this methodology, we further investigate the refraction of polaritons in biaxial vdW material and present its dependence on both $\alpha$ and $\theta$. The critical angles governing the transition of refraction between positive and negative can be analytically obtained via Eq. 2. To further demonstrate, we set $\alpha = 45°$ ($\theta = 90°$) to study the angle dependence on $\theta$ ($\alpha$) individually. For the case with fixed $\alpha = 45°$, four values of $\theta$ categorize the refraction of polaritons into positive refraction, negative refraction, and simultaneously positive and negative refraction. For the case with fixed $\theta = 90°$, negative refraction happens at arbitrary interface angles, while the refraction behavior of the polaritons varies within different ranges of $\alpha$. Lastly, we present tunable focusing in biaxial vdW materials where the focal point can be shifted both laterally and axially. Moreover, based on the electric tunability of materials such as BP, fixed focal length under different frequencies can be achieved. In conclusion, negative reflection, negative refraction, and electric tunability are achieved in biaxial vdW materials, showing promising potential in applications such as reconfigurable metalens[25] and transformation optics[22] for polaritons. In the future, we anticipate more exotic phenomena being observed in hybrid vdW systems, where vdW heterostructures and engineered substrates are introduced[29].

## METHODS

### Theoretical modeling

The permittivity tensor of α-MoO$_3$ is obtained by fitting the data in ref[26,28]. The isofrequency contour of the α-MoO$_3$ slab is obtained through the dispersion relation of anisotropic polaritons in the slab of biaxial crystals, see ref[39]. The surface conductivity tensor of a BP thin layer can be modeled by the effective conductivity, see ref[40] (see details in Section S7).

### Numerical simulations

Full-wave simulations of reflection and refraction are conducted via the commercial software *COMSOL Multiphysics 5.5*. The α-MoO$_3$ slab with a thickness of 200 nm is modeled to be embedded in air. The anisotropic polaritons are excited by a vertically oriented electric dipole placed 200 nm above the top surface of the α-MoO$_3$ slab at the center of the left boundary. Perfect matched layers are used surrounding the simulation region to prevent the reflection of the boundary. The field distributions in Figure 2e-g, Figure 3d-f, j-l, and Figure 4a-c are obtained at a plane of 20 nm above the top surface of the α-MoO$_3$ slab. The method to obtain the Fourier transform in Figure 3a-c, g-i is explained in Section S4.

BP thin layers are modeled as a surface embedded in the air with anisotropic conductivities shown in Eq. 6. The polaritons are excited by a *z*-oriented electric dipole placed 40 nm above the surface of the BP thin layer at the left boundary. In the incident medium, the imaginary part of surface conductivity along the y-axis is negative. On the transmitter side, the surface is rotated by 90°. Figure 4d-f are obtained by cutting an x-y plane 10 nm above the BP surface.

## ASSOCIATED CONTENT

**Supporting Information**

The following files are available free of charge.

Section S1, conventions of reflection and refraction diagram in *k*-space; Section S2, the plausibility of the choosing of the wavevectors; Section S3, the crystal direction ($\theta$) dependence of refraction; Section S4, the Fourier transform (FT) of the simulated field in Figure 3; Section S5, the interface direction ($\alpha$) dependence of refraction; Section S6, a comprehensive expression of Eq. 2; Section S7, theoretical modeling of biaxial vdW materials.

## AUTHOR INFORMATION

**Corresponding Author**

Cheng-Wei Qiu − Department of Electrical and Computer Engineering, National

University of Singapore, Singapore 117583, Singapore.

Email: eleqc@nus.edu.sg

**Author Contributions**

T.Z. and C.Z. contributed equally to this work. C.W.Q., T.Z., and C.Z. conceived the idea and designed the project. T.Z. and C.Z. carried the theoretical analysis and numerical simulations. All the authors have analyzed and discussed the results. T.Z. and C.Z. wrote the manuscript with inputs from Z.N.C and C.W.Q. C.W.Q. supervised the project.

**Notes**

The authors declare no competing interests.

## ACKNOWLEDGMENT

C.-W.Q. acknowledges financial support from the National Research Foundation, Prime Minister's Office, Singapore under Competitive Research Program Award NRF-CRP22-2019-0006. C.-W.Q. is also supported by a grant (R-261-518-004-720) from Advanced Research and Technology Innovation Centre (ARTIC).